\documentclass[onecolumn,aps,preprint,showpacs]{revtex4}
\usepackage{graphicx}

\begin{document}

\author{Yaroslav Tserkovnyak}
\affiliation{Lyman Laboratory of Physics, Harvard University, Cambridge, 
Massachusetts 02138}
\author{Arne Brataas}
\affiliation{Department of Physics, Norwegian University of Science and Technology, N-7491 Trondheim, Norway}
\author{Gerrit E. W. Bauer}
\affiliation{Department of NanoScience, Delft University of 
Technology, 2628 CJ Delft, The Netherlands}

\title{Dynamic exchange coupling and Gilbert damping in magnetic multilayers}

\begin{abstract}
We theoretically study dynamic properties of thin ferromagnetic films in contact with normal metals. Moving magnetizations cause a flow of spins into adjacent conductors, which relax by spin flip, scatter back into the ferromagnet, or are absorbed by another ferromagnet. Relaxation of spins outside the moving magnetization enhances the overall damping of the magnetization dynamics in accordance with the Gilbert phenomenology. Transfer of spins between different ferromagnets by these nonequilibrium spin currents leads to a long-ranged dynamic exchange interaction and novel collective excitation modes. Our predictions agree well with recent ferromagnetic-resonance experiments on ultrathin magnetic films.
\end{abstract}

\date{\today}
\maketitle


\section{Introduction}

The spin transfer in ferromagnet--normal-metal hybrids causes a number of exciting phenomena. Equilibrium spin currents across a thin normal-metal (\textit{N}) layer separating two ferromagnetic (\textit{F}) films explain the oscillating RKKY-type exchange coupling \cite{Sloncz:mmm93} as a function of the spacer thickness. This nonlocal exchange interaction may stabilize an antiparallel equilibrium configuration in magnetic multilayers \cite{Grunberg:prl86}, which display a giant magnetoresistance \cite {Baibich:prl88}. For thicker \textit{N} spacers or tunneling barriers, this static ferromagnetic exchange vanishes, but the \textit{F} layers can still be made to communicate by driving a dc current through the system. For example, depending on the direction of the current flow perpendicular to a layered \textit{F$\mid$N$\mid$F} spin-valve structure, a torque can be exerted on the magnetizations. When one magnetization is fixed, this torque favors either parallel or antiparallel configurations, and may lead to a switching of the other magnetization \cite{Sloncz:mmm96,Sloncz:mmm99}. Transport in the dynamic regime of moving magnetization directions has attracted relatively little attention, however, and is the subject of this paper.

We have recently launched the idea of spin pumping in weakly excited magnetic nanostructures. A moving ferromagnetic magnetization emits spins into adjacent conductors, exerting a relaxation torque and transferring an angular momentum out of the ferromagnet. The ejected spin angular momentum can either scatter back, relax in a nonmagnetic spacer, or be absorbed by a second ferromagnet. In the first case, the magnetization dynamics are not affected; in the second case, the magnetization motion is (nonlocally) damped by spin-flip scattering processes in the normal metal; and in the last case, the two ferromagnets become coupled by an exchange of itinerant spins, which may result in collective excitation modes. The spin-pumping concept has proven to be fruitful in the understanding of both the nonlocal damping mechanism \cite{Tserkovnyak:prl02} and dynamic exchange interaction \cite{Heinrich:prep} in hybrid \textit{F$\mid$N} systems. The functionality of magnetic devices, such as magnetic random-access memories, is strongly affected by relaxation characteristics of the magnetic media. 
The nonlocal relaxation and coupling can be large and even dominant compared with other mechanisms in ultrathin films, and can be engineered by composition, geometry, and magnetic configuration of hybrid \textit{F$\mid$N} systems. We hope that our theory will be useful as a designer tool, helping to realize future nanoscale magnetic devices with improved speed, operation threshold bias, and power consumption.

\section{Enhanced damping in single \textit{F} films}

Consider an \textit{F$\mid$N} bilayer as in Fig.~\ref{fig1}. Without a voltage bias, no spin or charge currents flow when the magnetization of the ferromagnet is constant in time. When the magnetization direction starts precessing (as, e.g., under the influence of an applied magnetic field), a spin current $\mathbf{I}_{s}^{\text{pump}}$ is pumped out of the ferromagnet into the \textit{N} layer \cite{Tserkovnyak:prl02}. When the ferromagnetic film is thicker than its transverse spin-coherence length $\lambda_{\text{sc}}$, $d>\lambda_{\text{sc}}=\pi/|k^{\uparrow}_{\text{F}}-k^{\downarrow}_{\text{F}}|$,
$k^{\uparrow(\downarrow)}_{\text{F}}$
being the spin-dependent Fermi wave vectors, this current depends on the complex-valued parameter $g^{\uparrow\downarrow}=g^{\uparrow\downarrow}_r+ig^{\uparrow\downarrow}_i$, the interfacial \textquotedblleft mixing\textquotedblright\ conductance \cite{Brataas:prl00,Waintal:prb00}, by 
\begin{equation}
\mathbf{I}_{s}^{\text{pump}}=\frac{\hbar }{4\pi }\left(g^{\uparrow\downarrow}_r\mathbf{m}
\times \frac{d\mathbf{m}}{dt}-g^{\uparrow\downarrow}_i\frac{d\mathbf{m}}{dt}\right) \,.
\label{Is}
\end{equation}
Here the time-dependent order parameter of the ferromagnet is a unit vector 
$\mathbf{m}(t)$, assuming a monodomain magnet with a spatially uniform magnetization at all times. A detailed derivation of Eq.~(\ref{Is}) based on the scattering-matrix theory is given in Refs.~\cite{Tserkovnyak:prl02,Tserkovnyak:prb022}. Alternatively, this result can be derived in the framework of magnetoelectronic circuit theory \cite{Brataas:prl00,Brataas:epjb01} using only energy and angular-momentum conservation \cite{Tserkovnyak:prb022}. The dc conductance matrix is defined by
\begin{equation}
g^{\sigma \sigma ^{\prime }}=\sum_{mn}\left[ \delta
_{mn}-r_{mn}^{\sigma }(r_{mn}^{\sigma ^{\prime }})^{\ast }\right]\,.   \label{g}
\end{equation}
Here $r_{mn}^{\uparrow }$ ($r_{mn}^{\downarrow }$) is the reflection coefficient off the \textit{F$\mid$N} interface for spin-up (spin-down) electrons on the normal-metal side, which can be obtained by \textit{ab initio} band-structure calculations \cite{Xia:prb02}.
$g^{\sigma \sigma ^{\prime }}$ defined in Eq.~(\ref{g}) must be renormalized for highly transparent interfaces, as explained in Ref.~\cite{Bauer:prep}, i.e., in the case of (globally) diffuse transport, the spurious Sharvin contributions have to be stripped off the real-valued conventional (Landauer-B\"{u}ttiker) resistances for a definite spin $\sigma$:
\begin{equation}
\frac{1}{\tilde{g}^{\sigma\sigma}}=\frac{1}{g^{\sigma\sigma}}-\frac{1}{2}\left(\frac{1}{N_N}+\frac{1}{N_{F\sigma}}\right)\,,
\label{gt1}
\end{equation}
while the inverse mixing conductance has to be corrected for the normal-metal Sharvin contribution only:
\begin{equation}
\frac{1}{\tilde{g}^{\uparrow\downarrow}}=\frac{1}{g^{\uparrow\downarrow}}-\frac{1}{2N_{N}}\,.
\label{gt2}
\end{equation}
The tildes in Eqs.~(\ref{gt1}) and (\ref{gt2}) denote the renormalized conductances, which reduce to the bare values (\ref{g}) when the numbers of transverse channels in the normal metal, $N_N$, and the ferromagnet, $N_{F\sigma}$, sufficiently exceed the contact conductances $g^{\sigma\sigma^\prime}$.

The total spin current, $\mathbf{I}_{s}$, across the \textit{F$\mid$N} interface also has a backflow contribution, $\mathbf{I}_{s}^{\text{back}}$, in addition to the pumped current, $\mathbf{I}_{s}^{\text{pump}}$, see Fig.~\ref{fig1}. The total spin transfer in the steady state,
\begin{equation}
\mathbf{I}_{s}=\mathbf{I}_{s}^{\text{pump}}-\mathbf{I}_{s}^{\text{back} }\,,
\label{sc}
\end{equation}
is determined self-consistently by the spin accumulation close to the interface. A finite value of the spin current, $\mathbf{I}_{s}$, would indicate the presence of a spin-sink mechanism in the normal metal by, e.g., spin-orbit coupling, when the spin angular momentum is transferred from the electron system to the lattice.
In the case of small spin relaxation, the dynamically created spin accumulation in the normal metal may serve as a spin-battery device \cite{Brataas:prb02}.
In the opposite spin-flip scattering regime, the angular-momentum loss of the ferromagnet by $\mathbf{I}_{s}$ results in a Gilbert-type damping of the magnetic precession \cite{Tserkovnyak:prl02}.

The spin current out of the ferromagnet carries angular momentum perpendicular to the magnetization direction, corresponding to a torque  $\text{\boldmath$\tau$}=-\mathbf{I}_{s}$ on the ferromagnetic condensate \cite{Sloncz:mmm96}. Disregarding interfacial spin flips, this torque is entirely transferred to the magnetization, which is described by a
generalized Landau-Lifshitz-Gilbert (LLG) equation \cite{Gilbert:pr55,Sloncz:mmm96} 
\begin{equation}
\frac{d\mathbf{m}}{dt}=-\gamma\mathbf{m}\times\mathbf{H}_{\text{eff}}
+\alpha_{0}\mathbf{m}\times\frac{d\mathbf{m}}{dt}+\frac{\gamma}{M_{s}V}
\mathbf{I}_{s}\,,  \label{llg}
\end{equation}
where $\gamma$ is the absolute value of the gyromagnetic ratio, $\alpha_{0}$ is the dimensionless intrinsic Gilbert-damping constant,
$M_{s}$ is the saturation magnetization of the ferromagnet, and $V$ is its volume. Referring to Eq.~(\ref{Is}) we see that the real part of the mixing conductance contributes to the damping just like the intrinsic bulk constant $\alpha_{0}$ which is thus smaller than the total Gilbert damping $\alpha=\alpha_{0}+\alpha^{\prime}$, whereas the imaginary part of the mixing conductance contributes like an effective field. The additional damping $\alpha^{\prime}$ is observable in, for example, ferromagnetic-resonance (FMR) spectra. Although not necessarily so for ferromagnetic insulators \cite{Huertas:prl02}, the mixing conductance for intermetallic \textit{F$\mid$N} interfaces is to a good approximation real \cite{Xia:prb02} and therefore
\begin{equation}
\alpha^\prime=\kappa\frac{\gamma\hbar\tilde{g}^{\uparrow\downarrow}}{4\pi M_sV}\,.
\label{ap}
\end{equation}
$\kappa=1$ corresponds to the perfect spin-sink model, when all pumped spins relax in the \textit{N} layer and the backflow $\mathbf{I}_{s}^{\text{back}}$ vanishes. $\kappa<1$ corresponds to a finite backflow \cite{Tserkovnyak:prb022}.

The spin-sink capacity of clean normal-metal layers in contact with a ferromagnetic film, as in Fig.~\ref{fig1}, is usually governed by spin-orbit scattering processes at impurities or defects. As the spin-flip probability $\epsilon=\tau_{\text{el}}/\tau_{\text{so}}$
(defined in terms of the elastic scattering time $\tau_{\text{el}}$ and spin-orbit relaxation time $\tau_{\text{so}}$)
rapidly increases with the atomic number $Z$,
$\epsilon\propto Z^4$ \cite{Abrikosov:zetf62,Meservey:prl78}, we expect a larger spin-sink effect [and therefore $\kappa$ in Eq.~(\ref{ap})] for heavier metals (impurities as well as hosts). The extent of the hybridization of the conduction bands with $p$ or $d$ orbitals also plays an important role. In particular, clean noble metals, Cu, Ag, and Au, with predominantly $s$ conduction electrons are poor spin sinks with correspondingly small $\epsilon$, but Pd and Pt, whose conduction electrons have significant $d$ character, have a high spin-orbit scattering rates and are efficient spin sinks. We note that heavy or magnetic impurities can turn an otherwise poor spin sink into a good one. The hierarchy of the Gilbert-damping enhancement has been measured for normal-metal buffers of Cu, Ta, Pd, and Pt (in order of increasing damping) \cite{Mizukami:jjap01}, in agreement with the aforementioned arguments \cite{Tserkovnyak:prb022}. In particular, the effect of spin pumping on the magnetization dynamics was shown to be negligible in the case of Cu, while Pt was a nearly perfect spin sink, resulting in a large increase of the Gilbert constant; this was also experimentally confirmed by Ingvarsson \textit{et~al.} \cite{Ingvarsson:prb02}.

To further investigate the spin-sink effect of the normal-metal buffer, we next consider a more complex system, consisting of a double \textit{N} layer attached to the ferromagnet, see Fig.~\ref{fig2}. An interesting situation arises when the layer \textit{N1} is a bad spin sink, such as Cu, and \textit{N2} is a perfect spin sink, such as Pt \cite{Tserkovnyak:prb022}. Mizukami \textit{et al.} \cite{Mizukami:mmm02,Mizukami:prb02} experimentally studied the FMR line width in permalloy (Py)$\mid$Cu$\mid$Pt composites as a function of Cu (\textit{N1}) width $L$. Next to $L$, there are three relevant length scales in the problem: the Fermi wave length, $\lambda_F$, the elastic scattering mean free path, $\lambda_{\text{el}}$, and the spin-diffusion length, $\lambda_{\text{sd}}$. In the coherent regime, $L<\lambda_F$, we do not expect large effects of a dusting \textit{N1} layer on the mixing conductance, which means that the damping enhancement (\ref{ap}) is approximately governed by $\tilde{g}^{\uparrow\downarrow}_{F\mid N2}$ (with a quantum modulation when $L\sim\lambda_F$).
If $L>\lambda_F$, the spin transport across the \textit{N1} spacer may be described by the diffusion equation \cite{Tserkovnyak:prb022}, provided that either the spacer or the interface is disordered.
In this regime, the spin backflow can be partitioned between the ferromagnet and layer \textit{N2}, see Fig.~\ref{fig2}. The relevant effective mixing conductance $\tilde{g}^{\uparrow\downarrow}_{\text{eff}}$ then has to account for scattering at both interfaces, \textit{F$\mid$N1} and \textit{N1$\mid$N2}, as well as in the \textit{N1} spacer:
\begin{equation}
\frac{1}{\tilde{g}_{\text{eff}}^{\uparrow\downarrow}}=
\frac{1}{\tilde{g}^{\uparrow\downarrow}_{F\mid N1}}+R_{N1}+
\frac{1}{\tilde{g}_{N1\mid N2}}\,,
\label{geff}
\end{equation}
where $R_{N1}$ is the resistance (per spin, in units of $h/e^2$) of the \textit{N1} layer, $\tilde{g}_{N1\mid N2}$ is the one-spin conductance of the \textit{N1$\mid$N2} interface, and we assumed that $L<\lambda_{\text{sd}}$. If in addition $L<\lambda_{\text{el}}$, the bulk scattering, $R_{N1}$, can be disregarded and the total resistance is given simply by the sum of the interfacial contributions. $\lambda_F$ thus sets the length scale for a sharp drop in the Gilbert-damping enhancement, followed by an algebraic decay with the effective mixing conductance (\ref{geff}) for $L>\lambda_F$. Such a damping drop was reported in Ref.~\cite{Mizukami:prb02} for Py$\mid$Cu$(L)$$\mid$Pt hybrids.

The regime $L>\lambda_F$ was studied in Refs.~\cite{Mizukami:mmm02,Mizukami:prb02}, where a smooth algebraic decay of the damping enhancement was measured for $L<\lambda_{\text{sd}}$ followed by an exponential suppression for thicker Cu spacers, in excellent agreement with our theory \cite{Tserkovnyak:prb022}. Mizukami et al. \cite{Mizukami:prb02} also offered an explanation for their measurements, using a phenomenological theory due to Silsbee \textit{et al.} \cite{Silsbee:prb79}. The theoretically calculated damping profiles, $\alpha(L)$, as a function of $L$ in Refs.~\cite{Tserkovnyak:prb022} and \cite{Mizukami:prb02} are barely distinguishable. In our opinion, the phenomenological approach employed in the latter has severe limitations, however. In particular, the separation between localized and conduction electron spins \cite{Silsbee:prb79,Mizukami:prb02} is not justified for itinerant ferromagnets like the transition metals. Furthermore, this separation leads to wrong results for insulating ferromagnets which, we believe, can generate a spin current into adjacent nonmagnetic conductors in the same way as a conducting ferromagnet \cite{Tserkovnyak:prb022}, in contrast to the prediction of Ref.~\cite{Silsbee:prb79}. The good fit of the experimental $\alpha(L)$ in Ref.~\cite{Mizukami:prb02} reflects the diffuse nature of transport in the copper layer rather than the spin-injection mechanism at the \textit{F$\mid$N} interface. The magnitude of the spin current as calculated in Ref.~\cite{Mizukami:prb02} is governed by a fitting parameter ($\Gamma$ in their notation) which cannot be verified or predicted by \textit{ab initio} calculations, unlike our spin-pumping expression (\ref{Is}).

The generality of our approach is demonstrated in the following by showing its validity for a qualitatively different system of two magnetic layers separated by a normal-metal film with negligible spin flip. We find that the same mechanism (namely the spin pumping) is responsible for the damping enhancement in coupled ferromagnets as well as the \textit{F$\mid$N} structures discussed so far.

\section{Dynamic coupling in spin valves}

In \textit{F$\mid$N$\mid$F} spin valves, the spin pumping causes qualitatively different effects in addition to the ones just described. Consider a system shown in Fig.~\ref{fig3}. In the following we take $L>\lambda_F$, so that quantum coherence (in particular the static exchange coupling) can be disregarded, but $L<\lambda_{\text{el}}$, so that the spins of electrons are transferred between the two ferromagnets ballistically.

The total spin current pumped into the normal metal,
\begin{equation}
\mathbf{I}_s^{\text{pump}}=\frac{\hbar}{4\pi}\left(\tilde{g}_1^{\uparrow\downarrow}\mathbf{m}_1\times\frac{d\mathbf{m}_1}{dt}+\tilde{g}_2^{\uparrow\downarrow}\mathbf{m}_2\times\frac{d\mathbf{m}_2}{dt}\right)\,,
\label{Ip}
\end{equation}
has contributions from both \textit{Fi$\mid$N} interfaces, see Eq.~(\ref{Is}). Here the small imaginary part of the mixing conductance \cite{Xia:prb02} is again disregarded. In a collinear configuration, the spin accumulation $\text{\boldmath$\mu$}_s$ induced by the spin pumping is perpendicular to the magnetizations.
This simplifies the expression for the bias-driven spin current to \cite{Brataas:epjb01,Tserkovnyak:prb022,Brataas:prb02}
\begin{equation}
\mathbf{I}_s^{\text{back}}=\frac{\tilde{g}_1^{\uparrow\downarrow}+\tilde{g}_2^{\uparrow\downarrow}}{4\pi}\text{\boldmath$\mu$}_s\,.
\label{I0}
\end{equation}
A steady state is established when the two spin currents, Eqs.~(\ref{Ip}) and (\ref{I0}), cancel each other:
$\mathbf{I}_s^{\text{back}}=\mathbf{I}_s^{\text{pump}}$.
The spin accumulation is then given by
$\text{\boldmath$\mu$}_s=4\pi(\tilde{g}_1^{\uparrow\downarrow}+\tilde{g}_2^{\uparrow\downarrow})^{-1}\mathbf{I}_s^{\text{pump}}$.
In general, $\mathbf{I}_{s2}=-\mathbf{I}_{s1}$, by the conservation of angular momentum, but the net spin current 
$\mathbf{I}_{sj}=\mathbf{I}_{sj}^{\text{pump}}-\mathbf{I}_{sj}^{\text{back}}$ across a single interface does not vanish. For the left ($j=1$) interface we find
\begin{equation}
\mathbf{I}_{s1} = \frac{\hbar}{4\pi}\frac{\tilde{g}_1^{\uparrow\downarrow}\tilde{g}_2^{\uparrow\downarrow}}{\tilde{g}_1^{\uparrow\downarrow}+\tilde{g}_2^{\uparrow\downarrow}}\left(\mathbf{m}_1\times\frac{d\mathbf{m}_1}{dt}-\mathbf{m}_2\times\frac{d\mathbf{m}_2}{dt}\right)\,.
\label{Is1}
\end{equation}
The normal-metal layer in our model scrambles the incoming spin current and divides it back over both ferromagnets. [For very transparent interfaces the scrambling is only partial, but the present treatment is still adequate if we properly renormalize the conductance parameters as in Eqs.~(\ref{gt1}) and (\ref{gt2}).] The second ferromagnet can thus cause damping of the precession in the first magnetic film, and \textit{vice versa}. A ferromagnetic layer can therefore serve as an efficient spin sink, just like a normal metal with strong spin-flip scattering.

We first discuss the implications of the spin transfer (\ref{Is1}) when the second ferromagnet, \textit{F2},
is stationary, $\mathbf{\dot{m}}_2=0$. In the FMR measurements, this is the case when one ferromagnet is in resonance, whereas the resonance frequency of the second ferromagnet is sufficiently different because of different magnetic anisotropies. The spin torque $\text{\boldmath$\tau$}=-\mathbf{I}_{s1}$ on the first ferromagnet (\textit{F1}) then has the form of the Gilbert damping when added as a source term to the LLG equation (\ref{llg}). In this case, the dynamic coupling of the ferromagnetic layers simply leads to an enhancement of the Gilbert-damping parameter with respect to its intrinsic value, exactly like in the case of single \textit{F} films. The damping enhancement, following from Eq.~(\ref{Is1}), of the \textit{F1} layer therefore reads
\begin{equation}
\alpha^\prime=\frac{\gamma\hbar\tilde{g}_1^{\uparrow\downarrow}\tilde{g}_2^{\uparrow\downarrow}}{4\pi M_sV(\tilde{g}_1^{\uparrow\downarrow}+\tilde{g}_2^{\uparrow\downarrow})}
\label{ap2}
\end{equation}
(with $\gamma$, $M_s$, and $V$ of \textit{F1}). Eq.~(\ref{ap2}) satisfactorily explains the increased viscous damping observed in Fe$\mid$Au$\mid$Fe spin valves \cite{Urban:prl01}. The physical nature of the damping in single \textit{F} films, Eq.~(\ref{ap}), versus spin valves, Eq.~(\ref{ap2}), is quite clear. The angular momentum is first driven out of the ferromagnet via spin pumping. In the former case (single ferromagnet), the subsequent spin-orbit processes in the normal metal relax injected spins and, as a result, slow down the coherent motion of the ferromagnet. In the latter case (spin valve), the spin angular momentum is transferred from ferromagnet \textit{F1} and into \textit{F2}. This spin current is absorbed by asynchronously driving the magnetization dynamics of \textit{F2}. Therefore, both the normal-metal and ferromagnetic spin sinks act as an external brake to slow down the precession of the resonantly excited magnet. The spin-sink efficiency of the normal metal is characterized by the spin-flip probability $\epsilon$ and its thickness $L$ \cite{Tserkovnyak:prb022}, which is lumped into the parameter $\kappa$ in Eq.~(\ref{ap}). It follows from Eq.~(\ref{ap2}) that the efficiency of the adjacent ferromagnet \textit{F2}, on the other hand, only depends on the \textit{F2$\mid$N} mixing conductance (as long as the second ferromagnet is thicker than its transverse spin-coherence length $\lambda_{\text{sc}}$).

In the following we consider dynamics of the coupled \textit{F1$\mid$N$\mid$F2}
system when both magnetizations are allowed to precess, i.e., when the ferromagnet resonance conditions are close to each other. By augmenting the LLG equation (\ref{llg}) for $\mathbf{m}_1$ by the spin current source term (\ref{Is1}) and with small variables $\mathbf{u}_i=\mathbf{m}_i-\mathbf{h}_i$,
where $|\mathbf{u}_i|\ll1$ and $\mathbf{u}_i\perp\mathbf{h}_j$, we obtain the linearized expression (assuming a circular precession)
\begin{equation}
\mathbf{\dot{u}}_1=\omega_1\mathbf{h}_1\times\mathbf{u}_1+\alpha_0\mathbf{h}_1\times\mathbf{\dot{u}}_1+\alpha_1^\prime\left(\mathbf{h}_1\times\mathbf{\dot{u}}_1-\mathbf{h}_2\times\mathbf{\dot{u}}_2\right)\,,
\label{u1}
\end{equation}
where $\omega_i$ is the resonance frequency of ferromagnet
\textit{Fi}, $\mathbf{h}_i$ is the unit vector in the direction of its effective magnetic field, and $\alpha_i^\prime=\gamma_i\hbar\tilde{g}_1^{\uparrow\downarrow}\tilde{g}_2^{\uparrow\downarrow}/[4\pi M_{si}V_i(\tilde{g}_1^{\uparrow\downarrow}+\tilde{g}_2^{\uparrow\downarrow})]$, subscript $i$ labeling corresponding quantities of \textit{Fi}. The dynamics of the magnetization direction $\mathbf{m}_2$ is obtained by exchanging subscripts $1\leftrightarrow2$ in Eq.~(\ref{u1}).

For a spin valve in the parallel configuration, $\mathbf{h}_i=\mathbf{h}$, and identical resonance frequencies, $\omega_i=\omega$, the quantity
$\mathbf{u}=\mathbf{u}_1/\alpha_1^\prime+\mathbf{u}_2/\alpha_2^\prime$ (which, up to a scaling factor, is the rf component of the total angular momentum) is affected only by the intrinsic bulk damping,
$\mathbf{\dot{u}}=\omega\mathbf{h}\times\mathbf{u}+\alpha_0\mathbf{h}\times\mathbf{\dot{u}}$, while the difference $\Delta\mathbf{u}=\mathbf{u}_1-\mathbf{u}_2$ relaxes according to $\Delta\mathbf{\dot{u}}=\omega\mathbf{h}\times\Delta\mathbf{u}+\alpha\mathbf{h}\times\Delta\mathbf{\dot{u}}$ with the enhanced (Gilbert) damping constant $\alpha=\alpha_0+\alpha_1^\prime+\alpha_2^\prime$.
The dynamic coupling in the antiparallel configuration as well as in the parallel configuration when the resonance frequencies have a large mismatch, $\Delta\omega\gg(\alpha_0+\alpha_1^\prime+\alpha_2^\prime)\omega$, favors relaxation of each magnetization towards the equilibrium configuration. In this case,
the two modes corresponding to excitations of either ferromagnet decouple,
each having an enhanced damping parameter $\alpha_i=\alpha_0+\alpha_i^\prime$.
This picture explains in detail the FMR profiles measured on the Fe$\mid$Au$\mid$Fe spin valves, not only in both symmetric and very asymmetric limits discussed herein but also in the intermediate regime of closely matched but different resonance frequencies in the parallel alignment \cite{Heinrich:prep}.

\section{Conclusions}

We presented a simple yet (for diffuse systems) complete theory of spin-current emission into adjacent conductors, which is generated by the time-dependent ferromagnetic order parameters. In the case of a single magnetic film in contact with a normal-metal layer, spin-flip processes in the latter viscously slow down the coherent precession of the ferromagnet, in accordance with the Gilbert phenomenology. When a magnetic layer is connected to other ferromagnets via Ohmic contacts, the latter can act as a nonlocal brake by relaxing the transverse angular momentum of the incident spins. Our framework formulated in terms of spin pumps and spin sinks successfully explains ferromagnetic resonance experiments on various multilayer systems.

In addition, this approach predicts a dynamic exchange coupling between ferromagnetic films connected by normal metals. Precessing magnetizations feel each other through a metallic spacer by exchanging nonequilibrium spin currents. Components of the ferromagnetic magnetizations that precess incoherently (i.e., out of phase) experience an additional damping obeying the Gilbert phenomenology. In symmetric spin valves and periodic superlattices, the dynamic exchange coupling causes a weakly damped, synchronized collective motion of the magnetic order parameters.

\acknowledgments

The authors are grateful to B.~I.~Halperin and B.~Heinrich for stimulating
discussions. This work was supported in part by the DARPA Award
No. MDA 972-01-1-0024, FOM, the NEDO International Joint Research Grant
Program \textquotedblleft Nano-magnetoelectronics\textquotedblright ,
and the NSF Grant No. DMR 99-81283.

\newpage

\begin{figure}[ph]
\caption{Schematic view of the \textit{F$\mid$N} bilayer. Precession of the magnetization direction $\mathbf{m}(t)$ of the ferromagnet \textit{F} pumps spins into the adjacent normal-metal layer \textit{N} by inducing a spin current $\mathbf{I}_{s}^{\text{pump}}$. This leads to a buildup of the normal-metal spin accumulation which either relaxes by spin-flip scattering or flows back into the ferromagnet as $\mathbf{I}_{s}^{\text{back}}$. The \textit{N} layer here is not an ideal reservoir but rather a film of the same cross section as the magnetic layer \textit{F}; the spin accumulation is position ($x$)-dependent.}
\label{fig1}
\end{figure}

\begin{figure}[ph]
\caption{Same as Fig.~\ref{fig1}, but now the normal-metal system is
composed of a bilayer \textit{N1$\mid$N2}. Ferromagnetic precession pumps spins
into the first normal-metal layer \textit{N1}. The spin buildup in
\textit{N1} may flow back into the ferromagnet \textit{F} as spin current
$\mathbf{I}_{s1}^{\text{back}}$, relax in \textit{N1}, or flow into the second
normal-metal layer \textit{N2} as spin current $\mathbf{I}_{s2}^{\text{back}}
$. The spin accumulation in \textit{N2} is disregarded since the layer is
assumed to be a perfect spin sink.}
\label{fig2}
\end{figure}

\begin{figure}[ph]
\caption{\label{fig3}Two ferromagnetic films \textit{F1} and
\textit{F2} connected by a normal-metal spacer \textit{N} of width $L$.
When the
magnetization directions $\mathbf{m}_1$ and $\mathbf{m}_2$ precess,
spin currents $\mathbf{I}_{s1}^{\text{pump}}$ and
$\mathbf{I}_{s2}^{\text{pump}}$ are pumped into the normal
metal. The spin-dependent chemical-potential imbalance
in \textit{N} causes the backflow of spin currents
$\mathbf{I}_{s1}^{\text{back}}$ and $\mathbf{I}_{s2}^{\text{back}}$.}
\end{figure}

\newpage

\includegraphics[scale=0.7,clip=]{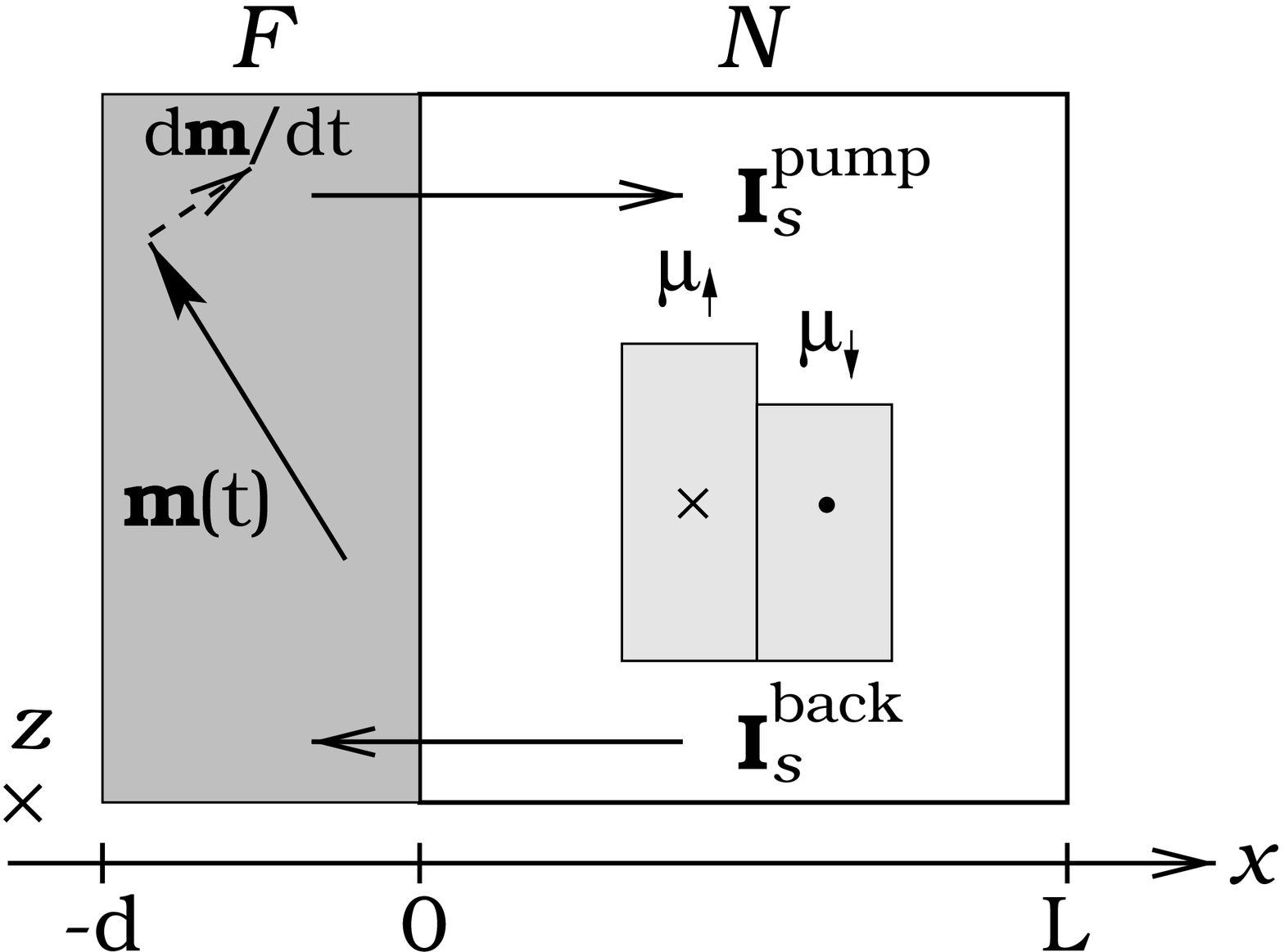}

\vspace{2cm}

{\Large Figure 1. Y. Tserkovnyak \textit{et al.}, J. Appl. Phys.}

\newpage

\includegraphics[scale=0.65,clip=]{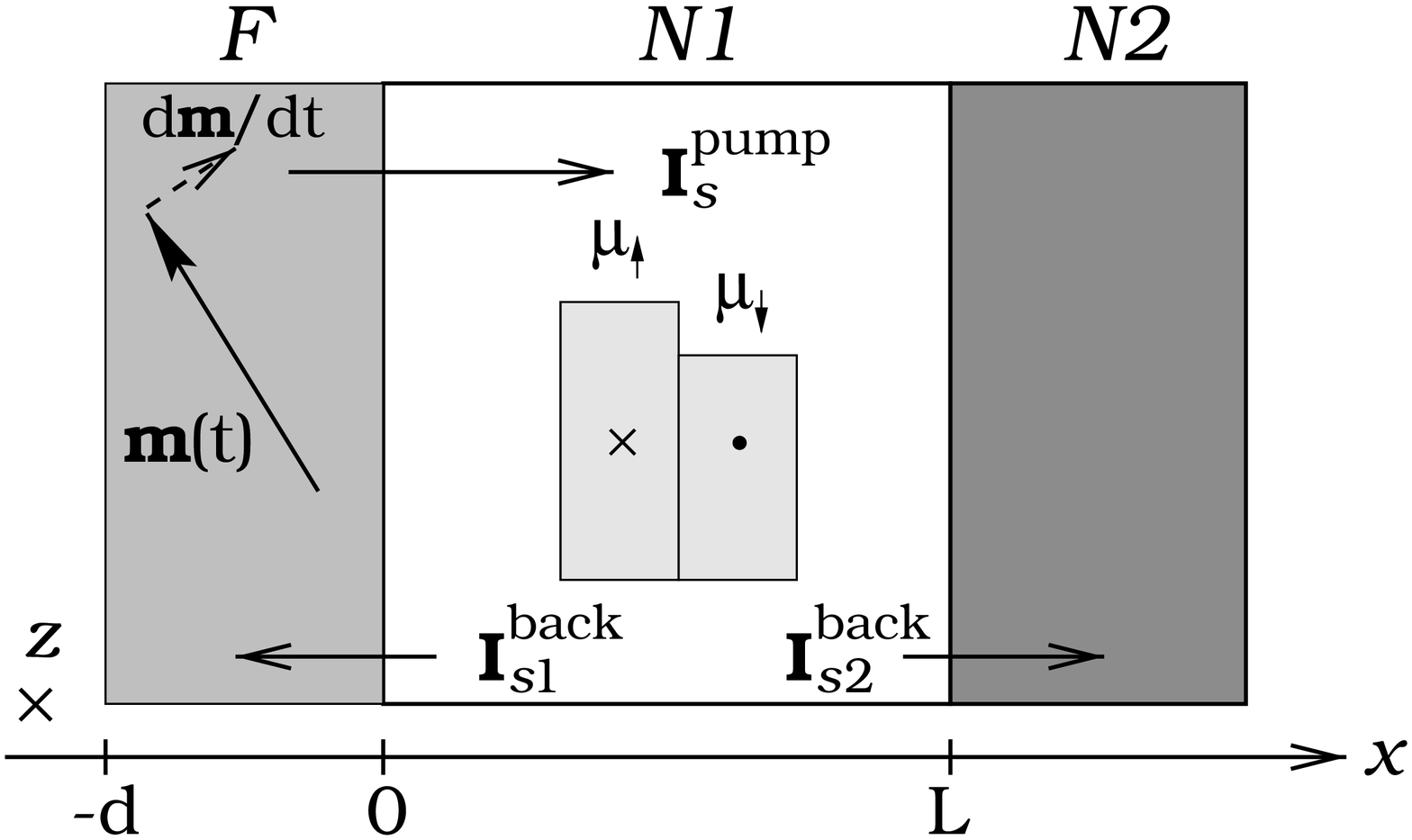}

\vspace{2cm}

{\Large Figure 2. Y. Tserkovnyak \textit{et al.}, J. Appl. Phys.}

\newpage

\includegraphics[scale=0.6,clip=]{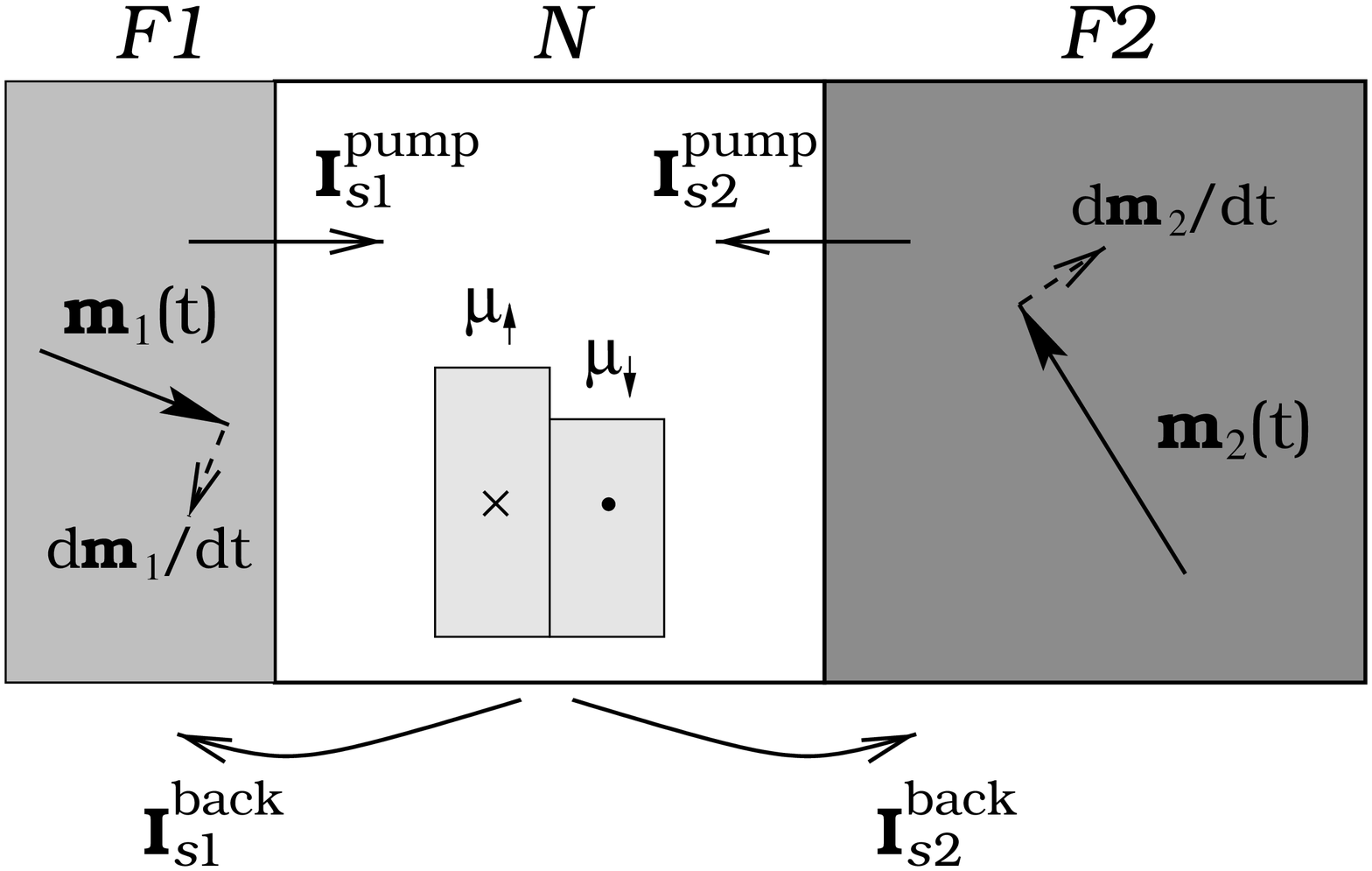}

\vspace{2cm}

{\Large Figure 3. Y. Tserkovnyak \textit{et al.}, J. Appl. Phys.}

\end{document}